\documentclass[a4paper,twoside]{article}
\usepackage{amsmath,amssymb,amsthm}
\usepackage{pstricks}
\usepackage{graphicx}

\newcommand{\C}{\mathbb{C}}
\newcommand{\R}{\mathbb{R}}
\newcommand{\Q}{\mathbb{Q}}
\newcommand{\CC}{\mathbb{C}}
\newcommand{\N}{\mathbb{N}}
\newcommand{\NN}{\mathbb{N}}
\newcommand{\Zi}{\mathbb{Z}}

\newcommand{\la}{\lambda}

\newtheorem{prop}{Proposition}[section]
\newtheorem{lem}[prop]{Lemma}
\newtheorem{cor}[prop]{Corollary}
\newtheorem{thm}{Theorem}[section]
\newtheorem{de}[prop]{Definition}

\begin{document}
%
%
\title{\textbf{Dynkin operators and renormalization group actions in pQFT}}
\author{Fr\'ed\'eric Patras\\
Laboratoire J.-A. Dieudonn\'e, CNRS UMR 6621,\\
Universit\'e de Nice,\\
Parc Valrose, 06108 Nice Cedex 02, France.
}%
\maketitle
\begin{abstract}
Renormalization techniques in perturbative quantum field theory were known, from their inception, to have a strong combinatorial content emphasized, among others, by Zimmermann's celebrated forest formula.
The present article reports on recent advances on the subject, featuring the role played by the Dynkin operators (actually their extension to the Hopf algebraic setting) at two crucial levels of renormalization, namely the Bogolioubov recursion and the renormalization group (RG) equations. For that purpose, an iterated integrals toy model is introduced to emphasize how the operators appear naturally in the setting of renormalization group analysis. The toy model, in spite of its simplicity, captures many key features of recent approaches to RG equations in pQFT, including the construction of a universal Galois group for quantum field theories.
\end{abstract}

\section*{Introduction}
Renormalization techniques in perturbative quantum field theory (pQFT) were known, from their inception, to have a strong combinatorial content emphasized, among others, by the celebrated Zimmermann's forest formula.
A. Connes and D. Kreimer and, more recently, A. Connes and M. Marcolli, have brought forward that one should try to understand the physics underlying these phenomena by means of tools of noncommutative geometry. Their understanding that a Hopf algebra structure on Feynman graphs underlies the forest formula, their extension to prounipotent groups of the classical Birkhoff decomposition of loops and their discovery of a universal Galois group underlying renormalization are three milestones for our current mathematical understanding of the theory \cite{ck1998,ck2000,ck2001,cm2004,cm22004,cm2006}.

However useful, deep and possibly leading to a new understanding of fundamental interactions, this noncommutative picture has certain drawbacks. For example, a key ingredient of the noncommutative approach is the hypothesis that the local geometry of the fundamental interactions around the space-time dimension, as encoded in the dimensional regularization (DR) process, governs the physics. How far this assumption is valid is unclear (at least to the author of the present article). Indeed, although DR has particularly nice properties (e.g. with respect to the various gauge symmetries of quantum field theories), all the other usual regularization processes and renormalization schemes that have apparently nothing to do with the geometrical phenomena arising around the space-time dimension (Pauli-Villars, zero-momentum subtraction, cut-offs, and so on) lead ultimately to the same results \cite{iz}. This suggests that, besides geometry, classical combinatorics and algebra should also contribute in an essential way to the renewal of the mathematical foundations of pQFT.

The purpose of the present article is to report on advances related to this research program, emphasizing the ubiquity of the Dynkin operator, especially from the point of view of the renormalization group (RG) analysis of regularized and renormalized amplitudes. 
To make clear the relevance of the operator for the RG analysis, we introduce a simple toy model for RG actions, namely by studying how solutions of linear differential equations are changed when a rescaling of the equation (reminiscent of the rescaling of energy scales in the usual RG analysis of pQFT) is performed. As we shall see, the corresponding RG equations encode \it exactly \rm our Dynkin operator analysis of the beta function in pQFT \cite{egp2007}. We strongly believe that, because of this phenomenon, because of the ubiquity of the Dynkin operator, and because of the relative simplicity and tractability of the underlying algebra, this approach can be meaningful beyond the classical setting of pQFT.

The article is organized as follows. The first section shows how the classical Dynkin operator shows up naturally when performing a RG analysis of solutions of linear differential equations. The second one recalls briefly some general facts about Hopf algebras and renormalization in pQFT, emphasizing the Bogoliubov recursion. The purpose of the third one is twofold. We show how the constructions in the first section lift to the algebraic setting of graded connected Hopf algebras. In the process we recall the construction of a universal Galois group acting on arbitrary graded Hopf algebras. Letting this group act on the particular class of Hopf algebras of Feynman graphs, one recovers the Connes-Marcolli universal Galois group of pQFT. The fourth section emphasizes the role of the Dynkin operator as a universal operator for RG analysis. As an example, we show how the Bogoliubov recursion can be understood that way. The last one focuses on the RG analysis of regularized and renormalized amplitudes in pQFT. Here, the striking result is that the beta function (the generator of the RG equation, where the RG action is obtained by a rescaling of the mass scale) is best understood by a RG analysis of the counterterm (where the RG action on the counterterm is driven now by the number of loops in Feynman diagrams and not any more by a rescaling of mass scales).

This article is largely based on joint works with Ch. Reutenauer \cite{patreu1999,patreu2002,patreu22002} and K. Ebrahimi-Fard, J. Gracia-Bond\`\i a and D. Manchon \cite{egp2007,egp22007,emp2008,emp22008}. It has benefited from many discussions with J.-Y. Thibon and owes a tribute to the theory of noncommutative symmetric functions as developed in \cite{gelfand1995}. Precise references to these works will be given in the article. A warm thanks goes to them for fruitful ongoing collaborations and particularly to K. Ebrahimi-Fard and J. Gracia-Bond\`\i a for several comments on a first version of this article.

\section{RG analysis of differential equations}

Let us consider the evolution equation: 
$$A'(t)=A(t)H(t)$$
with initial condition $A(0)=1$, where $A(t)$ and $H(t)$ belong to a matrix algebra or, more generally, to an operator algebra. The solution is given by the time-ordered exponential series:
$$T(\exp \int\limits_0^tH(x)dx):=1+\int\limits_0^tdt_1H(t_1)+...+\int\limits_{\Delta_t^n}dt_1\dots dt_nH(t_1)\dots H(t_n)+\dots$$
where $\Delta_t^n:=\{0\leq t_1\leq \dots\leq t_n\leq t\}$. 

The general idea behind RG analysis is to study the dependency of a physical system on changes of scales. To put the emphasis on this dependency, we will call systematically (and most probably slightly abusively) the corresponding differential equation the RG equation. In the present situation, one of the simplest possible changes of energy scales consists in rescaling the operator $H(t)$ to $H^\lambda(t):=\lambda H(t)$, with $\lambda\in\R^\ast$.
Writing $A^\lambda(t)$ for the solution of ${A^\lambda}'(t)=A^\lambda(t)H^\lambda(t)$, we get $A^\lambda(t)=T(\exp \int\limits_0^tH^\la(x)dx)$.
The problem becomes algebraically interesting and nontrivial when we try to derive a renormalization group equation, that is a differential equation solved by $A^\lambda(t)$ with respect to the perturbation parameter $\la$. As we shall see later, the computation parallels exactly the one for the RG equations solved by regularized and renormalized amplitudes in pQFT.

In order to derive this differential equation, let us recall a few elementary facts on permutations and iterated integrals. For further details on descents in symmetric groups, we refer to \cite{reutenauer1993} and \cite{BlessenohlS}. The more algebraically oriented reader may also refer to the Hopf algebraic approach in \cite{patras1994} that will underly most of our forthcoming constructions.

\begin{de}
The descent set of a permutation $\sigma\in S_n$ is the subset of $[n-1]$ defined by:
$$Desc(\sigma ):=\{i<n,\sigma (i)>\sigma(i+1)\}.$$
\end{de}

With the notation $\sigma =(\sigma(1),\dots , \sigma(n))$, we get, for example:
$Desc((42315))=\{1,3\}.$
In general, to each  subset $S$ of $[n-1]$ are associated two corresponding ``Solomon elements'' in $\Q [S_n]$:
$$D_{=S}^{(n)}:=\sum\limits_{\sigma\in S_n, Desc(\sigma)=S}\sigma,\ D_{S}^{(n)}:=\sum\limits_{\sigma\in S_n, Desc(\sigma)\subseteq S}\sigma.$$
When no confusion can arise, we write simply $D_{=S}$ and $D_S$ for $D_{=S}^{(n)}$ and $D_{S}^{(n)}$.
By inclusion/exclusion (i.e. M\"obius inversion), the two are related by the formulas:
$$D_S=\sum\limits_{T\subseteq S}D_{=T},\ D_{=S}=\sum\limits_{T\subseteq S}(-1)^{|S|-|T|}D_{T}.$$

With the notation 
$$H_\sigma:=[\int\limits_{\Delta_t^n}dt_1\dots dt_nH(t_1)\dots H(t_n)]\cdot\sigma :=
\int\limits_{\Delta_t^n}dt_1\dots dt_nH(t_{\sigma(1)})\dots H(t_{\sigma(n)})$$
and the extension of the notation to elements $\beta\in \C[S_n], \beta=\sum_{\sigma\in S_n}\mu_\sigma\cdot \sigma$,
$$H_\beta:=[\int\limits_{\Delta_t^n}dt_1\dots dt_nH(t_1)\dots H(t_n)]\cdot\beta :=
\sum_{\sigma\in S_n}\mu_\sigma[\int\limits_{\Delta_t^n}dt_1\dots dt_nH(t_1)\dots H(t_n)]\cdot \sigma ,$$
the product of $k$ iterated integrals reads:
$$H_{n_1}H_{n_2}...H_{n_k}=H_{n_1+...+n_k}\cdot D_{\{n_1,n_1+n_2,...,n_1+...+n_{k-1}\}}^{(n_1+...+n_k)}=H_{D_{\{n_1,n_1+n_2,...,n_1+...+n_{k-1}\}}^{(n_1+...+n_k)}}$$
where we have written $H_{p}$ for $H_{Id_p},\ Id_p=(1...p)\in S_p$. We omit the proof that follows easily by splitting the integration domain into simplices. The construction underlies the classical Mielnik-Pleba\'nski \cite{MP70} explicit solutions to the continuous and discrete Baker-Campbell-Hausdorff formulas by means of Solomon's eulerian idempotent (the logarithm of the identity in the convolution algebra of the symmetric groups group algebras, see \cite{solomon1968,reutenauer1986,patras1992,patras1993,reutenauer1993}).
More generally, we have:
$$H_{D_{S_1}^{(n_1)}}H_{D_{S_2}^{(n_2)}}=H_{D_{S_1\coprod \{n_1\}\coprod S_2+n_1}^{(n_1+n_2)}},\ \ H_{D_{=S_1}^{(n_1)}}H_{D_{=S_2}^{(n_2)}}=H_{D_{=S_1\coprod S_2+n_1}^{(n_1+n_2)}}+H_{D_{=S_1\coprod \{n_1\}\coprod S_2+n_1}^{(n_1+n_2)}}$$
and:
$$H_{D_{S_1}^{(n_1)}}H_{D_{S_2}^{(n_2)}}...H_{D_{S_k}^{(n_k)}}=H_{D_{S_1\coprod \{n_1\}\coprod S_2+n_1\coprod \{n_1+n_2\} .... \coprod \{n_1+...+n_{k-1}\}\coprod S_k+n_1+...+n_{k-1}}^{(n_1+...+n_k)}}.$$

Let us recall now the definition of the Dynkin operator. From the point of view of Lie theory, this is the element ${\cal D}_n$ of $\Zi[S_n]$ such that, for an arbitrary sequence of elements $x_1,...,x_n$ in an associative algebra $A$:
$$[\dots [x_1,x_2],...,x_n]=x_1....x_n\cdot {\cal D}_n$$
where $[x,y]=xy-yx$ is the Lie bracket on $A$ and where the symmetric group acts on the right: $x_1....x_n\cdot\sigma :=x_{\sigma(1)}...x_{\sigma(n)}\ \forall\sigma\in S_n$. For example, ${\cal D}_2=(12)-(21)=D_\emptyset-D_{=\{1\}}\in\Zi[S_2]$ and ${\cal D}_3=(123)-(213)-(312)+(321)=D_\emptyset-D_{=\{1\}}+D_{=\{12\}}$ since
$[x_1,x_2]=x_1x_2-x_2x_1,\ [[x_1,x_2],x_3]=x_1x_2x_3-x_2x_1x_3-x_3x_1x_2+x_3x_2x_1$.
In general, an easy recursion shows that ${\cal D}_n=\sum\limits_{i=0}^{n-1}(-1)^iD_{=\{1,...,i\}}$.

\begin{thm}\label{toyRG}
The RG equation for $A^\la (t)$ reads:
$$\lambda\cdot{\partial\over{\partial \la}}A^\la (t)=A^\la (t)\cdot \beta(t)$$
where the beta function (the infinitesimal generator of the RG equation) is given by $\beta(t):=\sum\limits_{n=1}^\infty H^\la_{n}\cdot {\cal D}_n$.
\end{thm}

Indeed, let us expand the left and right-hand sides of the equation into homogeneous components with respect to the powers of $\la$. The component of degree $n$ in the left hand side is simply ${n}H^\la_{n}(t)$. 
In the right-hand side, we get:
$$\sum\limits_{i=0}^{n-1} H^\la_{i}\cdot (H^\la_{n-i}\cdot {\cal D}_{n-i}).$$
Expanding this sum according to the product rule for iterated integrals, we get:

$\sum\limits_{i=0}^{n-1}\sum\limits_{j=0}^{n-i-1}(-1)^jH_i^\lambda\cdot H_{D_{=\{1,...,j\}}}^\la$
$$=\sum\limits_{j=0}^{n-1}(-1)^j H_{D_{=\{1,...,j\}}}+
\sum\limits_{i=1}^{n-1}\sum_{j=0}^{n-i-1}(-1)^jH^\la_{D_{=\{i,i+1,...,i+j\}}}+\sum\limits_{i=1}^{n-1}\sum_{j=0}^{n-i-1}(-1)^jH^\la_{D_{=\{i+1,...,i+j\}}}$$
$$=H_{D_n^\la}+\sum\limits_{j=1}^{n-1}(-1)^j H_{D_{=\{1,...,j\}}}+\sum\limits_{j=0}^{n-2}(-1)^j H_{D_{=\{1,...,1+j\}}}$$
$$+\sum\limits_{i=2}^{n-1}\sum_{j=0}^{n-i-1}(-1)^jH^\la_{D_{=\{i,i+1,...,i+j\}}}+\sum\limits_{i=1}^{n-1}H_{D_n}^\la+
\sum\limits_{i=1}^{n-2}\sum_{j=1}^{n-i-1}(-1)^jH^\la_{D_{=\{i+1,...,i+j\}}}$$
$$=n\cdot H^\la_{n}$$
from which the Theorem follows.

A striking corollary of the RG equation for $A^\la (t)$ is that $A^\la(t)$ and the corresponding beta function $\beta(t)$ satisfy a formula similar to the one for the universal singular frame in the noncommutative geometrical and motivic Galoisian approach to pQFT \cite{cm2004}. A similar identity also holds in the setting of noncommutative symmetric functions \cite{gelfand1995}. This is not a mere coincidence, since identities involving Solomon elements in descent algebras translate into properties of Hopf algebras \cite{patras1994} and noncommutative symmetric functions \cite{gelfand1995}, and since Hopf algebras provide the right algebraic framework to understand the combinatorial structures of pQFT \cite{ck2000,ck2001} -we postpone an explanation of these phenomena to the forthcoming sections of the article.

\begin{cor}\label{dyninv}
Let us write $A^\la (t)=\sum\limits_nA_n^\la (t)$ and $\beta( t)=\sum\limits_n\beta_n(t)$, where $A_n^\la(t)$ and $\beta_n(t)$ refer to the homogeneous component of degree $n$ (with respect to the parameter $\la$) in $A^\la(t)$ and $\beta(t)$. We have:
$$A_n^\la(t)=\sum\limits_{l\leq n}\sum\limits_{k_1+...+k_l=n,k_i>0}\frac{\beta_{k_1}(t)...\beta_{k_l}(t)}{k_1(k_1+k_2)...(k_1+...+k_l)}$$
\end{cor}

Indeed, since $kA_k^\la(t)=\sum\limits_{i=1}^kA_{k-i}^\la(t)\beta_i(t)$,
we have $A_n^\la(t)=\frac{\beta_n(t)}{n}+\sum\limits_{i=1}^{n-1}\sum\limits_{1\leq j\leq n-1}\frac{A_{n-i-j}^\la(t)\beta_j(t)\beta_i(t)}{(n-i)n}.$
Substituting iteratively the value of $A_k^\la(t)$ in the right hand side of the identity, we arrive at the required identity:
$$A_n^\la(t)=\sum\limits_{l\leq n}\sum\limits_{k_1+...+k_l=n,k_i>0}\frac{\beta_{k_1}(t)...\beta_{k_l}(t)}{k_1(k_1+k_2)...(k_1+...+k_l)}.$$

\section{Hopf algebras and RG analysis in QFT}

The present, short, section will serve as a reminder of the algebraic structures underlying the renormalization process and as a motivation for the constructions in the next sections. There are several good accounts of the subject, both classical (such as Collins' \cite{collins1984}) and recent (such as Manchon's \cite{man2001} and Figueroa and J.~M.~Gracia-Bond\'{\i}a's \cite{fg2005}). For that reason, we omit details and limit our account to the leading ideas. Readers familiar with the subject should skip it and proceed directly to the next section. The others should hopefully get insights into the logic behind the renormalization process and some intuition why the algebraic constructions that will be developed later make sense for pQFT.

The main problem of renormalization is to associate a finite quantity to the divergent integrals associated by certain rules (the Feynman rules) to Feynman diagrams (precisely one-particle-irreducible, superficially divergent Feynman diagrams). Using these quantities,  physically relevant quantities (such as scattering amplitudes) may be computed through the LSZ reduction formulas \cite{iz}. Concretely, the Feynman diagrams are build out of vertices with prescribed numbers of edges. The vertices represent the fundamental interactions of the theory, whereas the edges stand essentially for the propagation of particles between two interactions. The diagrams arise from a perturbative expansion, and the orders of the expansion correspond naturally to the number of loops in the diagrams.

At first order these quantities (the amplitudes) look like:
$$\phi (m)=\int_{\R^4}\frac{d^4q}{(q^2+m^2)^2}$$
and divergences show up.
Various techniques have been deviced to remove this kind of  divergences. 
The process is usually a twofold one. First, one introduces one (or several) non physical regularization parameter(s) (regularization step), and the integrals are expanded according to this parameter. An asymptotic study then reveals the structure of the divergences; their removal produces finite quantities (subtraction step). Renormalization group equations arise from the study of the dependence of the solutions in this/these extra parameter(s). Together with the derivation of finite quantities for a given value of the parameter, the RG equations are at the core of the renormalization process.

In practice, there are several techniques (cut-off, Pauli-Villars, Taylor subtraction, dimensional regularization (DR),...). 
Most of them give satisfactory results (see e.g. \cite{iz}) but (excepted maybe for the cut-off techniques), their physical grounds are still not plainly understood. One of the hopes raised by the noncommutative geometrical approach to renormalization was precisely to pave the way for a new understanding of these phenomena. Although some ideas presented below hold for other schemes than DR (this was proved explicitly for Taylor subtraction in \cite{egp2007}), we will focus on DR which is certainly the theory most easily dealt with from the mathematical point of view, besides presenting many nice features such as compatibility with gauge invariance.

In the DR process, a regularization parameter $\epsilon$ is introduced that should be understood as an infinitesimal deformation of the space-time dimension:
$$4\rightsquigarrow D=4-\varepsilon$$so that $d^4q$ is changed to $d^Dq$. 
A typical dimensional regularization computation then reads, for $\phi (m)$ as above \cite{zeidler}, $$\phi^{reg}(m;\epsilon,\mu):=i\mu^\epsilon\int_{\R^{4-\epsilon}}\frac{d^{4-\epsilon}q}{(q^2+m^2)^2}=i\pi^2\Gamma(\frac\epsilon 2)(\frac{m\pi^{\frac 1 2}}\mu)^{-\epsilon}$$
$$=i\pi^2(\frac 2\epsilon -{\mathcal C}-2\ln (\frac{m\pi^{\frac{1}{2}}}{\mu}))+o(1)$$
where $\mathcal C$ stands for Euler's constant and $o(1)$ refers to $\epsilon\longmapsto 0$.
A simple subtraction step (or ``minimal subtraction'', another example is treated in \cite{zeidler}) consists in removing the divergence. The substracted divergent part $\phi_{-}(m;\epsilon,\mu):=-i\pi^2(\frac 2\epsilon)$ is called the counterterm. We will refer to the remaining quantity $\phi_{+}(m;\epsilon,\mu) :=  i\pi^2( -{\mathcal C}-2\ln (\frac{m\pi^{\frac{1}{2}}}{\mu}))+o(1)$ and to its constant part $\phi^{ren}(m;\epsilon,\mu):=\phi_{+}(m;0,\mu)$ as the renormalized amplitudes.
A concrete example (the Ginzburg–-Landau $\phi_4^4$
scalar model in Euclidean field theory) is treated explicitly in \cite{fg2005}, to which we refer, also for precise explanations on notions such as superficial degree of divergence or subdivergences of a Feynman diagram, which we omit to explain but are essential to the understanding of these phenomena.
An important point to notice is that, due to the particular structure of DR, an extra ``mass scale'' parameter $\mu$ is needed in the regularization process to insure that the physical dimension of the regularized amplitude remains correct. In DR, the RG analysis is conducted according to this mass scale parameter.

The process we considered (regularization, asymptotic expansion, removal of the divergence by minimal subtraction) describes correctly what happens at the first order of the expansion (the one-loop order). At higher orders, the regularization principle remains unchanged, but the removal of the singular part (arising from the removal of divergences in the asymptotic expansion) is made more complex. Roughly, some subdiagrams $\gamma$ of a given Feynman $\Gamma$ diagram may contribute directly to make the overall integral associated to $\Gamma$ divergent. These subdivergences have to be removed recursively in order to make the overall renormalization process physically consistent. The recursion is conducted according to the partial order of (sub)diagrams inclusions. This is the purpose of the Bogoliubov recursion, which is best explained in terms of a Hopf algebra structure on diagrams.

This Hopf algebra $H$ is defined as the polynomial algebra generated by the empty set $\emptyset$ and
the connected proper (or 1PI) Feynman graphs that are (superficially) divergent and/or have (superficially)
divergent subdiagrams, with set union as the
product operation. The empty set is therefore the unit. The product is written $\pi$. Since $H$ is freely generated as a (graded) commutative algebra by connected proper diagrams (a diagram with $n$ loops is attributed the degree $n$), the coproduct is well defined once it is defined on connected proper diagrams $\Gamma$. We get:
$$\Delta (\Gamma ) :=\sum\limits_{\emptyset\subseteq \gamma\subseteq\Gamma}\gamma\otimes \Gamma/\gamma.$$
The sum is over all divergent, proper, not necessarily connected subdiagrams of $\Gamma$. 
The map $\Delta$ is coassociative and an algebra map from $H$ to $H\otimes H$, making $H$ a true Hopf algebra (the existence of a unit, of a counit and of an antipode follows immediately from the graded structure of the product and the coproduct on $H$). The counit (which maps $\emptyset$ to 1  and the nontrivial graphs to 0 is written $\eta$). We refer to \cite{fg2005} for the necessary refinements (e.g. how tadpoles should be handled).

We are now in the position to introduce the setting of the group-theoretic approach to renormalization. Since bare Feynman rules lead to divergences, we have to deal with
regularized Feynman rules $\phi^{reg}$. Such a rule defines a map from the set $\cal F$ of (connected, proper) Feynman diagrams to the algebra of Laurent series $A:={\bf C}[\epsilon^{-1},\epsilon ]]$ and extends uniquely, since $H$ is a polynomial algebra over $\cal F$,
to a $A$-valued character of $H$, that is, a multiplicative map from $H$ to ${\bf C}[\epsilon^{-1},\epsilon ]]$.
Equivalently, $\phi^{reg}$ it is an element of the prounipotent affine group $G_{{\bf C}[\epsilon^{-1},\epsilon ]]}(H)$ of $A$-valued characters on $H$, where the group structure is induced by the convolution of linear maps from $H$ to ${\bf C}[\epsilon^{-1},\epsilon ]]$:
$$f\ast g:H\stackrel{\Delta}{\rightarrow} (H\otimes H)\stackrel{f\otimes g}{\rightarrow}({\bf C}[\epsilon^{-1},\epsilon ]]\otimes {\bf C}[\epsilon^{-1},\epsilon ]])\stackrel{\pi}{\rightarrow} {\bf C}[\epsilon^{-1},\epsilon ]],$$
where $\pi$ stands for the product map.

The Bogoliubov recursion is a process allowing the inductive
construction of a decomposition
$$\phi^{reg}=\phi_-^{-1}\ast \phi_+,$$
where $\phi_-\in G_-(A)$ and $\phi_+\in G_+(A)$ are respectively the counterterm character and the renormalized character (they formalize the notion of counterterm and renormalized amplitude, as introduced previously).
Here, $G_-(A)$ (resp. $G_+(A)$) stands for the group of algebra maps $\gamma$ from $H$ to $\C[\epsilon^{-1}]$ such that, for any Feynman diagram $\Gamma$, $\gamma(\Gamma)\in A^-:=\epsilon^{-1}\C[\epsilon^{-1}]$ (resp. the group of algebra maps from $H$ to $A^+:=\C[[\epsilon]]$).
Here, the crucial point, first noticed by Ch. Brouder and D. Kreimer \cite[Eq.(55)]{krei99} is that the projection map $R$ of Laurent series on their divergent part (orthogonally to ${\bf C}[[\epsilon]]$):
$$R:A= {\bf C}[\epsilon^{-1},\epsilon ]]\longrightarrow A^-$$
satisfies the relation:
$$ R (x y) + R(x) R(y) = R (R(x) y + x R(y)) .$$
That is, $R$ defines a weight one Rota-Baxter algebra structure on $A$ \cite{rota69,rotasmith72}. This relevance of Rota-Baxter algebra structures for pQFT was first noticed by K. Ebrahimi-Fard and brought into full light in his Thesis (and in a series of papers) \cite{e2006}, stimulating many recent works. Many ideas developed in this article do actually hold for Rota-Baxter algebras of an arbitrary weight $\theta$, that is for algebras provided with an operator $R$ such that the following identity holds:
$$ \theta R (x y) + R(x) R(y) = R (R(x) y + x R(y)) .$$ 
We refer e.g. to \cite{eg2006,emp2008}, also for further historical references on the subject.

The Bogolioubov recursion now reads:
$$\phi_-=-R(\overline{\phi}),\ \phi_+=(1-R)(\overline{\phi})$$
where $\overline\phi$ is Bogoliubov's preparation map:
$$\overline{\phi}=\phi_-\ast (\phi^{reg} -\eta )$$
Equivalently, $\phi^{reg} =\phi_-^{-1}\ast\phi_+$ and $\phi_-$ solves:
$$\phi_-=1-R(\phi_-\ast (\phi^{reg} -\eta ))$$
The recursion process is by induction on the number of loops in diagrams. The fact that the recursion
defines elements of $G_-(A)$ and $G_+(A)$ is not obvious from the
definition and follows from the multiplicativity constraints, as was
shown by Kreimer and Connes-Kreimer, see e.g.
\cite{krei99,ck2000,em2007} and the references therein, also for complementary insights on the Bogoliubov recursion -such as its solution by means of the so-called BCH recursion.

The remaining part of the article will be focused on showing how \it both \rm the Bogoliubov recursion and the RG equations (the study of the dependency of $\phi_+$ and $\phi_-$ on the mass scale parameter) can be understood by means of the Dynkin operator, and how our RG equation for solutions of linear differential equations generalizes in this setting.

\section{A universal Galois group for Hopf algebras and pQFT}

The purpose of the present section is to show how to construct a universal Galois group acting on Hopf algebras. The construction goes back to \cite{patras1994} and is encoded in the notion of the ``descent algebra of a Hopf algebra''. 

Notice immediately that, since Feynman diagrams are organized into a Hopf algebra, the construction of a universal Galois group acting on Hopf algebras incorporates as a particular case the notion of a universal Galois group for pQFT, as suggested by P. Cartier and constructed by A. Connes and M. Marcolli.
This was first noticed in \cite{egp2007}. This observation has a further interest: indeed, the structure of the universal Galois group of the theory of Hopf algebras carries a very rich structure. Among others, besides a graded Hopf algebra structure on its algebra of coordinates ${\cal D}esc^\ast$ (the graded dual of the descent algebra, to be introduced below), one can show that each of the graded components of ${\cal D}esc$ carries another associative product that has nice compatibility properties with respect to the Hopf algebra structure \cite{patras1994,gelfand1995,patreu2002}. One can expect these extra-structure to be meaningful for pQFT.

Let us formalize first the computations on iterated integrals in the first section of the article. For a standard approach to the descent algebra, we refer to \cite{reutenauer1993}.

\begin{de}
Let us write ${\cal D}esc$ for the linear span of Solomon elements $D_S^{(n)}$, where $n$ runs over $\N$ and $S$ over the subsets of $[n-1]$. The
vector space ${\cal D}esc$ is provided uniquely with a cocommutative Hopf algebra structure by the following requirements:
\begin{itemize}
\item The product, written $\ast$ on ${\cal D}esc$ is given by:
$$D_S^{(n)}\ast D_T^{(m)}:=D_{S\coprod \{n\}\coprod T+n}^{(n+m)}$$
\item The coproduct $\Delta$ is defined by requiring that the $D_\emptyset^{(n)}$ form a sequence of divided powers, that is,
$$\Delta (D_\emptyset^{(n)}):=\sum\limits_{i=0}^nD_\emptyset^{(i)}\otimes D_\emptyset^{(n-i)}$$
\end{itemize}
\end{de}

A fundamental Theorem, due to Solomon, shows that the graded components ${{\cal D}esc}_n\subset\Q[S_n]$ of ${\cal D}esc$ are subalgebras of $\Q[S_n]$ (where the product is induced by the product of permutations in $S_n$). The algebra ${{\cal D}esc}_n$ is usually referred to as Solomon's algebra of type $A_n$ (see \cite{solomon1976}).

The choice of the product $\ast$ is dictated by the product formulas for iterated integrals.
The reasons for the choice of the coproduct are explained in \cite{patreu2002}. They insure that the Hopf algebra structure on ${\cal D}esc$ is compatible with the action on Hopf algebras, as described in the next Theorem.

\begin{de}
The universal Galois group $Gal_{Hopf}$ of Hopf algebra theory is the set of group-like elements in the Hopf algebra ${\cal D}esc$, that is:
$$Gal_{Hopf}:=\{x\in {{\cal D}esc}, \Delta(x)=x\otimes x\}$$
\end{de}

This terminology (which is not standard) is chosen here to emphasize the connexion with the (canonically isomorphic) universal Galois group  of renormalization (or motivic Galois group) of \cite{cm2004}.

\begin{thm}\label{ggroup}

\ 

Let $H=\bigoplus\limits_{n\in{\bf N}}H_n$ be an arbitrary graded connected commutative (resp. cocommutative) Hopf algebra. The descent algebra ${\mathcal D}esc$ and the universal Galois group $Gal_{Hopf}$ act naturally on $H$. Moreover, the Solomon algebra of type $A_n$ (resp. the opposite algebra) acts naturally on $H_n$.
\end{thm}

The construction of the action is fairly straightforward: as an obvious corollary of the definition of the product $\ast$, the descent algebra is a free associative algebra over the $D_{\emptyset}^{(n)}$. Let us define the \it descent algebra ${\mathcal D}esc_H$ of a Hopf algebra $H$ \rm as the convolution subalgebra of $End(H)$ generated by the graded projections $p_n:H\longmapsto H_n$ (orthogonally to $\bigoplus\limits_{m\not= n}H_m$). Here, the definition of the convolution product in $End(H)$ mimics the definition of the convolution of linear morphisms on $H$ as introduced in the previous section and is omitted.
 Then, since ${\mathcal D}esc$ is free associative, there is a unique algebra map from ${\mathcal D}esc$ to ${\mathcal D}esc_H\subset End(H)$. This induces a natural action of ${\mathcal D}esc$ on $H$.

The tricky point is the verification that the various structures existing on the descent algebra (such as the product on ${{\cal D}esc}_n$ or the coproduct) are compatible in a natural way with this action. For these compatibilities, we refer to \cite{patras1994,patreu2002}. 

A nice corollary of this construction, that we mention for the sake of completeness, is that the structure theorems for Hopf algebras (Leray, Cartier-Milnor-Moore) can be recovered in a transparent combinatorial way from this approach. We refer to \cite{patras1992,patras1993,patras1994} for details. As could be expected already from Solomon's analysis of the Poincar\'e-Birkhoff-Witt Theorem, the Solomon's eulerian idempotents play a key role in this construction.

\begin{cor} The universal Galois group of Hopf algebra theory is a combinatorial, explicit, realization of the universal Galois group of renormalization theory.
\end{cor}

The fact that the universal Galois group of Hopf algebra theory is a universal group for perturbative quantum field theories is a direct consequence of the Thm~\ref{ggroup} and of the existence of a Hopf algebra structure governing the combinatorics of Feynman diagrams.

\section{The Dynkin operator as a universal operator for the RG analysis}

A corollary of the universal properties of the descent algebra is that our toy-model RG equation for iterated integrals (Thm.~\ref{toyRG}) can be lifted to the Hopf algebraic setting. This lift is based on the study of Dynkin operators actions on Hopf algebras in \cite{patreu2002} and is the main purpose of the present section. This is illustrated with the example of the Bogoliubov recursion, following \cite{egp22007,emp2008}.

So, let us turn back for a while to the iterated integral approach to RG analysis. One characteristic feature of the time-ordered exponential expansion of the solution of a first-order linear differential equation is that the $n$-th order term of the expansion (an $n$-dimensional iterated integral) rescales as $\lambda^n$ under the action of the Renormalization Group. There is a universal feature to this property: very much as the descent algebra is a universal object for Hopf algebra theory, the Dynkin operator is a universal operator for RG analysis, at least for elementary renormalization groups such as $\CC^\ast$.

Let us consider first a formal power series in noncommuting variables and a scalar parameter $\lambda$, $S(\lambda)=1+\sum\limits_{n\in\NN^\ast}\lambda^ns_n$. Mimicking our computation with the time-ordered exponential, we define the series $B$ by:
$$\lambda S'(\lambda)=S(\lambda)B(\lambda).$$
The requirement $S(0)=1$ insures that $B(\la)$ is well-defined.
When the series $S(\la)$ originates from physics or analysis, e.g. as a solution of a differential equation, it is usually expected to have a group-like behavior. It is therefore natural to lift the computations to the algebraic setting, view the variables $s_n$ as a family of free noncommuting variables, and provide the free associative algebra $\mathcal F$ generated by the $s_n$ with a Hopf algebra structure by requiring that the coproduct $\Delta$ be such that:
$$\Delta (S(\lambda))=S(\lambda)\otimes S(\lambda).$$
This equation is enough to define the coproduct and the Hopf algebra structure (since it implies that $\Delta(s_n)=\sum\limits_{k\leq n}s_k\otimes s_{n-k}$). These definitions can be traced back to the seminal work \cite{gelfand1995}.

\begin{prop}
We have, in $\mathcal F$:
$$B(\lambda)={\mathcal D}(S(\lambda))$$
where ${\cal D}:=\sum\limits_{n\geq 1}{\cal D}_n$ stands for the sum of Dynkin operators acting naturally on the Hopf algebra $\mathcal F$  in view of Thm~\ref{ggroup}.
\end{prop}
 
Indeed, let us notice first that, since the RG equation of Thm.~\ref{toyRG} holds for an arbitrary time-dependent operator $H(t)$, it translates into an equation for the Dynkin operators in the descent algebra, namely:
$${\cal D}:=\sum\limits_{n\geq 1}{\cal D}_n=(\sum\limits_{n\in\N}D_\emptyset^{(n)})^{-1}\ast (\sum\limits_{n\in\N}n\cdot D_\emptyset^{(n)}).$$
This identity, usually derived from the presentation of the Dynkin operator as an iterated bracketing, can be traced back to von Waldenfels, see \cite{patreu2002} for details. The proposition follows, since $S(\la)$ is group-like and since $ D_\emptyset^{(n)}(S(\la))=\la^ns_n$, so that:
$$(\sum\limits_{n\in\N}D_\emptyset^{(n)})^{-1}\ast (\sum\limits_{n\in\N}n\cdot D_\emptyset^{(n)})(S(\la))=S(\la)^{-1}\la S'(\la).$$

Let us show now how the Bogoliubov construction of counterterms in pQFT can be accounted for with this construction.
Let $A$ be an associative algebra provided with a linear endomorphism $L$. For $x\in A$, let us consider the series $X(\lambda)$ solution of the equation:
$$X(\lambda)=1+L(X(\lambda)\cdot \lambda x)$$
with initial condition $X(0)=1$, so that the equation can be solved recursively as: $X(\lambda)=1+\lambda L(x)+\lambda^2 L(L(x)x)+...$. When $X$ is the solution of the Bogoliubov recursion in pQFT, $L$ is the projection operator $R$ on the divergent part of Laurent series whereas $-x$ is the regularized amplitude $\phi^{reg} -\eta$.
The RG analysis can then be performed as in $\mathcal F$ (since any identity satisfied by a formal power series with free variables holds automatically for an arbitrary power series with coefficients in a associative algebra), and we get the RG equation:
$$\lambda X'(\lambda)=X(\lambda)Y(\lambda)$$
where $Y(\lambda)$ is the image in $A$ of the corresponding element $B(\lambda)$ in $\mathcal F$ (where $s_1$ is sent to $L(x)$, $s_2$ to $L(L(x)x)$, and so on). 

The way this general pattern specializes to the Bogoliubov recursion is particularly intriguing, since it leads to new formulas and ideas for the construction of the counterterm in pQFT which have an independent mathematical interest. The corresponding analysis was performed in \cite{egp22007,emp2008} in the more general framework of arbitrary Rota-Baxter algebras and, although we state them in the context of the Bogoliubov recursion for dimensionally regularized amplitudes in pQFT, the following results hold in that general setting.

Recall first the notion of (left) pre-Lie or Vinberg algebra:
this is a vector space equipped with a bilinear product $\bullet$ such that:
$$(a\bullet b)\bullet c-a\bullet (b\bullet c)=(b\bullet a)\bullet c-b\bullet (a\bullet c)$$
so that, in particular, the bracket $[a,b]_\bullet$ satisfies the Jacobi identity.
The algebra $\mathcal L$ of maps from the Connes-Kreimer Hopf algebra of Feynman graphs to the algebra of Laurent series carries a pre-Lie structure, defined by:
$$a\bullet b:=R(a)b-bR(a)+ba,$$
where $R$ stands as usual for the projection on the divergent part of the Laurent series.
We set: for $a\in \mathcal L$, 
$$l^{(n)}(a):=l^{(n-1)}(a)\bullet a$$
$$r^{(n)}(a):=R(l^{(n)}(a)).$$

\begin{prop} The solution $\phi_-^\la$ to the generalized Bogolioubov recursion $\phi_-^\la =1+R(\phi_-^\la\ast\la(\phi^{reg}-\eta))$ satisfies the RG equation:
$$\la{\phi_-^\la }'=\phi_-^\la\cdot \xi(\la)$$
where $\xi(\la)=\sum\limits_{n\geq 1}r^{(n)}(\la(\phi^{reg}-\eta))$.
\end{prop}

The computation of $\xi(\la)$ is far from obvious and depends heavily on the existence of a Rota-Baxter algebra structure on the algebra of Laurent series. We refer to \cite{egp22007,emp2008} for a proof. The physical relevance of this equation is largely unclear at the moment since the rescaling $\phi^{reg}\longmapsto \lambda\phi^{reg}$ (which breaks the multiplicative properties of the regularized amplitude $\phi^{reg}$) does not seem to have any obvious physical meaning. However, as we already observed, the mathematical structures underlying the formula are very rich, interesting on their own, and suggest a relevance of pre-Lie structures in the construction of counterterms and renormalized amplitudes.

\section{RG equations and the beta function in pQFT}

Whether or not the previous analysis can be extended to include the classical RG equations in pQFT may seem problematic. Indeed, in that case, the RG analysis has to take into account the phenomenon of renormalization, so that understanding algebraically the RG equations requires the understanding of the interactions between the behavior of divergences in regularized amplitudes and the changes of mass scales.
However, surprisingly, the Dynkin operator approach to RG equations still holds in that case, leading to a new, Lie theoretic, understanding of the beta function. 

Let $H$ be, once again, the graded connected commutative Hopf algebra of Feynman diagrams of a given quantum field theory treated perturbatively. As usual, $A={\mathbf C}[\varepsilon^{-1},\varepsilon ]]$ stands for the algebra of Laurent series.
To the (pro-unipotent) group of characters $G(A)$ (the group of algebra morphisms from $H$ to $A$)
is naturally associated a (pro-nilpotent) Lie algebra (its elements are called infinitesimal characters):
$L(A):=\{\phi\in Hom(H^+,A), (H^+)^2\subset Ker(\phi )\}$
where $H^+:=\bigoplus\limits_{n>0}H_n$. 
Since $G$ and $L$ are pro-unipotent and pro-nilpotent, they behave essentially as a group of unipotent (resp. a Lie algebra of nilpotent) matrices and the $log$ and the $exp$ maps induce bijections between $G$ and $L$. However, the logarithm and exponential maps are not suited to the understanding of the RG equations: as could be expected from our previous analysis of RG equations, one has to use instead of the classical logarithm the Dynkin operator:

\begin{thm}\label{logder}

Right composition with $\cal D$ is a
bijective map from $G(A)$ to $L(A)$. The inverse map is given by
$$
    \Gamma : \alpha \in L(A) \longmapsto \sum \limits_n
    \sum_{{k_1,\dots,k_l\in \mathbb{N}^\ast \atop k_1 + \dots +
    k_l=n}} \frac{\alpha_{k_1} \ast
    \dots \ast \alpha_{k_n}} {k_1(k_1+k_2) \dots (k_1+ \dots +k_n)} \in G(A).
$$
\end{thm}

The computation of the inverse follows from the identity in the Cor.~\ref{dyninv}, when suitably translated into an identity in the descent algebra -and, therefore, into a universal identity for Hopf algebras. A purely algebraic proof was first obtained in \cite{egp2007}, to which we refer for details.

\

Recall that, in dimensional regularization, the regularized amplitude depends on $\varepsilon$ and on the mass scale $\mu$. The mass scale appears at the power $\mu^{n\varepsilon}$ in the regularized amplitude of an $n$-loop graph $\Gamma$ (we write $n=|\Gamma|$).
Here, the (mass scale) renormalization group (MRG) is the group of rescalings of $\mu$, and acts therefore on the group of $A$-valued characters $G(A)$ as
a {{one-parameter action}} of~$\mathbf{C}^* \ni t$:
$$
  \phi^t(\Gamma) :=
  t^{\varepsilon|\Gamma|}\phi(\Gamma),
  $$
where, for simplicity, we write from now on $\phi$ for $\phi^{reg}(m;\varepsilon, \mu)$ and $\Gamma$ stands for an arbitrary graph (not necessarily connected).
We have, for $\phi^t \in G(A)$ the decomposition:
$$
    \phi^t = {(\phi^t)}_-^{-1} \ast {(\phi^t)}_+.
$$
For reasons that will become clear soon, we also introduce a second renormalization group (the ``loop'' RG, LRG) action:
$$
  \phi^{(\la)}(\Gamma) :=
  \la^{|\Gamma|}\phi(\Gamma),
  $$
  which is the one we have encountered in the previous section and are now familiar with (according to which the components of degree $n$ in a Hopf algebra, here the graphs with $n$ loops, are rescaled by a factor $\la^n$). The two RG actions commute and, moreover, the LRG action commutes to the Bogoliubov decomposition in the sense that:
  $$
  (\phi^t)^{(\la)} = ((\phi^t)^{(\la)})_-^{-1} \ast ({(\phi^t)^{(\la)}})_+=({(\phi^t)}_-^{-1})^{(\la)} \ast ({(\phi^t)}_+)^{(\la)}
$$
Moreover,
$$t\cdot\frac{\partial}{\partial t}\phi^t=\epsilon\cdot\la\frac{\partial}{\partial\la}{(\phi^t)^{(\la)}}_{|\la=1}.$$
The physical constraints associated to the renormalization process translate then into the \it locality condition\rm :

\begin{thm} [Collins] Let $\phi$ be a
dimensionally regularized Feynman rule character. The counterterm
character in $\phi^{t}= (\phi^{t})^{-1}_-\ast(\phi^{t})_+$
satisfies
$$
    t\,\frac{\partial{(\phi^t)}_-}{\partial t} = 0
$$
\end{thm}

Or ${(\phi^t)}_-$ is equal to~$\phi_-$, i.e. independent of~$t$. The physical reason for this is that the counterterms can be taken mass-independent;
this goes back at least to \cite{ArcheoCollins}. For this fact, and more details on the physical significance
of the locality property in pQFT, we refer the reader to \cite{collins1984,ck2001}. 

We say the $A$-valued characters with this property are
{\it{local}} characters: $\phi \in G^{\rm{loc}}(A) \subset G(A)$. Notice that, by definition of the MRG action, $G_+(A)\subset G^{loc}(A)$. We will be mainly interested in local counterterms, that is, elements in $G_-(A)^{loc}:=G_-(A)\cap G^{loc}(A)$. Notice that $G_-(A)^{loc}$ is stable under the LRG action. Performing a RG analysis for this action yields immediately a differential equation for the counterterm:

\begin{lem}
We have, for any $\psi\in G_-(A)$, the LRG equation
$$\lambda ({\psi^{(\lambda)}})'=\psi^{(\lambda)}\cdot (\psi^{(\la)}\circ\cal D).$$
Moreover, the map $\psi\longmapsto \psi\circ \cal D$ is a bijection between $G_-(A)$ and $L(A_-)$.
\end{lem}

We omit the proof since it follows, once again, from the universal properties of the Dynkin operator for RG analysis and from the existence of a Hopf algebra structure on $H$.

Actually, one can show \cite{egp2007} that the map $\psi \mapsto Dyn(\psi ):=
\varepsilon(\psi\circ {\cal D})$ is a bijection between $G^{\rm loc}_-(A)$ and
$L(\mathbb{C})$. Interesting phenomena show up when one tries to understand locality from this point of view since, for local characters, the infinitesimal generators of the LRG equation (for the counterterm) and of the MRG equation (for the renormalized character) surprisingly agree:

\begin{thm} Let $\phi\in G^{loc}(A)$. For the renormalized character
$\phi_{\rm ren}(t):= (\phi^t)_+(\varepsilon=0)$, we have:
$$
    t\frac{\partial}{\partial t} \phi_{\rm ren}(t) = Dyn(\phi_-^{-1})
    \ast \phi_{\rm ren}(t),
$$
the abstract RG~equation\end{thm}

Indeed, since $\phi$ is local, we have: $$\phi^t=(\phi^t)^{-1}_-\ast (\phi^t)_+=\phi_-^{-1}\ast (\phi^t)_+.$$
and get:
$$
    t\frac{\partial(\phi^t)_+}{\partial t}= \phi_- \ast t\frac{\partial}{\partial t}\phi^t={\phi_-\ast \epsilon\la\frac{\partial}{\partial\la}(\phi^t)^{(\la)}}_{|\la =1}$$
    $$=\phi_-\ast \epsilon\la {(\frac{\partial}{\partial\la}(\phi_-^{-1})^{(\la)}\ast(\phi_+^t)^{(\la)}+(\phi_-^{-1})^{(\la)}\ast\frac{\partial}{\partial\la}(\phi_+^t)^{(\la)})}_{|\la=1}$$
    $$=Dyn(\phi_-^{-1})\ast\phi_+^t+{\epsilon\la\frac{\partial}{\partial\la}(\phi_+^t)^{(\la)}}_{|\la=1}.$$
    Taking the limit when $\epsilon\rightarrow 0$, we get:
   $$t\frac{\partial}{\partial t} \phi_{\rm ren}(t) = Dyn(\phi_-^{-1})\ast\phi_{ren}(t).$$

These equations show in the end that the classical abstract RG equation can be understood as an equation involving the action of the Dynkin operators of the Hopf algebras of Feynman graphs. Moreover, the simple toy model introduced at the begining of the article accounts fairly well for the combinatorial subtelties of the construction.
These simple arguments, that do not appeal to the complex structures of noncommutative geometry but rely on the more classical tools of free Lie calculus and symmetric group actions, appear to us in the end as a simple (down-to-earth, but efficient) way to account for the universal Galoisian approach to renormalization. Together with K. Ebrahimi-Fard and J. Gracia-Bond\`\i a (with whom this research program is currently conducted), we hope these ideas to be flexible enough to accomodate for the more complex computations showing up in ``real life'' pQFT (e.g. with higher dimensional renormalization groups corresponding to several coupling constants; with other renormalization schemes than RG, and so on).


\begin{thebibliography}{99}

\bibitem{BlessenohlS}
    D.~Blessenohl and M.~Schocker,
    Noncommutative character theory of the symmetric group,
    World Scientific, Singapore, 2005.

\bibitem{OldNiko89}
    N.~Bourbaki,
    Elements of Mathematics. Lie groups and Lie algebras. Chapters 1--3,
    Springer, Berlin, 1989.


\bibitem{ArcheoCollins}
    J.~C.~Collins,
    Structure of the counterterms in dimensional regularization,
    \textit{Nucl. Phys. B} 80 (1974) 341--348.

\bibitem{collins1984}
    J.~C.~Collins,
    Renormalization,
    Cambridge University Press, Cambridge, 1984.

\bibitem{ck1998}
    A.~Connes and D.~Kreimer,
    Hopf algebras, renormalization and noncommutative geometry,
    \textit{Commun. Math.~Phys.} 199 (1998) 203--242.

\bibitem{ck2000}
    A.~Connes and D.~Kreimer,
    Renormalization in quantum field theory and the Riemann--Hilbert
    problem I. The Hopf algebra structure of graphs and the main theorem,
    \textit{Commun. Math. Phys.} 210 (2000) 249--273.

\bibitem{ck2001}
    A.~Connes and D.~Kreimer,
    Renormalization in quantum field theory and the Riemann--Hilbert
    problem. II. The $\beta$-function, diffeomorphisms and the
    renormalization group,
    \textit{Commun. Math. Phys.} 216 (2001) 215--241.

\bibitem{cm2004}
   A.~Connes and M.~Marcolli,
   Renormalization and motivic Galois theory.
   \textit{Internat. Math. Res. Notices}
   Vol. 2004 76 (2004), 4073--4091.

\bibitem{cm22004}
    A.~Connes and M.~Marcolli,
    From Physics to Number Theory via Noncommutative Geometry II:
    Renormalization, the Riemann--Hilbert correspondence, and motivic
    Galois theory,
    to appear in~\textit{Frontiers in Number Theory, Physics and
    Geometry}.

\bibitem{cm2006}
    A.~Connes and M.~Marcolli,
    Quantum Fields and Motives,
    \textit{J.~Geom.~Phys.} 56 (2006) 55--85.
    
\bibitem{e2006}
    K. Ebrahimi-Fard,
    Rota–Baxter Algebras and the Hopf Algebra of Renormalization,
    Ph.D. Thesis, University of Bonn, 2006.
    
\bibitem{egp2007}
    K. Ebrahimi-Fard, J. Gracia-Bond\`\i a and F. Patras,
    A Lie theoretic approach to renormalization, 
    \textit{ Comm. Math. Phys} 276, (2007) 519–549.

\bibitem{egp22007}
    K. Ebrahimi-Fard, J. Gracia-Bond\`\i a and F. Patras,
    Rota-Baxter algebras and new identities, 
    \textit{Letters Math. Phys.} 81, (1), (2007), 61-75. 

\bibitem{eg2006}
    K.~Ebrahimi-Fard and L.~Guo,
    ``Rota--Baxter Algebras in Renormalization of Perturbative Quantum Field Theory'',
    to appear in \textit{Fields Institute Communications}. 
    
\bibitem{em2007}    
    K. Ebrahimi-Fard, D. Manchon,
    The combinatorics of Bogoliubov's recursion in renormalization,

\bibitem{emp2008}
    K. Ebrahimi-Fard, D. Manchon and F. Patras,
    A Bohnenblust-Spitzer identity for Rota-Baxter algebras solves Bogolioubov's counterterm recursion, 
    \textit{J. Noncommutative Geom.}
    To appear.

\bibitem{emp22008}
    K. Ebrahimi-Fard, D. Manchon and F. Patras,
    New identities in dendriform algebras, 
    \textit{J. Algebra.}
    320 (2),  (2008), 708-727.

\bibitem{fg2005}
    H.~Figueroa and J.~M.~Gracia-Bond\'{\i}a,
    Combinatorial Hopf algebras in quantum field theory I,
    \textit{Reviews of Mathematical Physics} 17 (2005) 881--976.

\bibitem{gelfand1995}
    I.~M.~Gelfand, D.~Krob, A.~Lascoux, B.~Leclerc, V.~Retakh and
    J.-Y.~Thibon,
    Noncommutative symmetric functions,
    \textit{Adv. Math.} 112 (1995) 218--348.


\bibitem{iz}
    C. Itzykson and J. Zuber,
    Quantum field theory,
    MacGraw-Hill, New York (1980)

\bibitem{krei99}
    D. Kreimer,
    Chen's Iterated Integral represents the Operator Product Expansion,
    Adv. Theor. Math. Phys. 3.3 (1999).

\bibitem{man2001} 
    D. Manchon,
    Bogota lectures on Hopf algebras, from basics to applications to renormalization, 
    Comptes-rendus des Rencontres math\'ematiques de Glanon 2001 (2003).
    
\bibitem{MP70}
    B. Mielnik and J. Pleba\'nski,
    Combinatorial approach to Baker-Campbell-Hausdorff exponents,
    \textit{Ann. Inst. Henri Poincar\'e, Section A}
    Vol.~XII, (1970), 215--254.   
    
\bibitem{patras1992}
    F.~Patras,
    \textit{Homoth\'eties simpliciales},
    Th\`ese de Doctorat, Paris 7, Janvier 1992.         

\bibitem{patras1993}
    F.~Patras,
    La d\'ecomposition en poids des alg\`ebres de Hopf,
    \textit{Ann.~Inst.~Fourier} 43 (1993) 1067--1087.

\bibitem{patras1994}
    F.~Patras,
    L'alg\`ebre des descentes d'une big\`ebre gradu\'ee,
    \textit{J. Algebra} 170 (1994) 547--566.
    
\bibitem{patreu1999}
    F. Patras and  Ch. Reutenauer,
    Higher Lie idempotents,
    \textit{J. Algebra}  222, (1999), 51-64.     

\bibitem{patreu2002}
    F.~Patras and Ch.~Reutenauer,
    On Dynkin and Klyachko idempotents in graded bialgebras,
    \textit{Adv. Appl. Math.} 28 (2002) 560--579.
    
\bibitem{patreu22002}
   F. Patras and  Ch. Reutenauer,
   Lie representations and an algebra containing Solomon's,
   \textit{J. Alg. Comb.}  16, (2002), 301-314. 
    
\bibitem{reutenauer1986}
    C. Reutenauer,
    Theorem of Poincar\'e-Birkhoff-Witt, logarithm and representations of the symmetric group whose orders are the Stirling numbers. 
    \textit{Combinatoire enum\'erative}, Proceedings, Montr\'eal (1985), (ed. G. Labelle and P. Leroux). Lecture Notes in Mathematics, 267--284, Springer, Berlin.
    
\bibitem{reutenauer1993}
    C.~Reutenauer,
    Free Lie algebras,
    Oxford University Press, Oxford, 1993.

\bibitem{rota69}
    G.-C.~Rota,
    ``Baxter algebras and combinatorial identities. I,II'',
    \textit{Bull.~Amer.~Math.~Soc.} {\bf75} (1969)~325--329;
    \textit{ibidem} {\bf75} (1969)~330--334.

\bibitem{rotasmith72}
    G.-C.~Rota and D.~A.~Smith,
    ``Fluctuation theory and Baxter algebras'',
    \textit{Symposia Mathematica} {\bf IX} (1972)~179--201.
    
\bibitem{solomon1968}
    L.~Solomon,
    On the Poincar\'e-Birkhoff-Witt theorem,
    \textit{J. Combinatorial Theory} 4 (1968) 363--375.
    
\bibitem{solomon1976}
    L.~Solomon,
    A Mackey formula in the group ring of a Coxeter group,
    \textit{J. Algebra} 41 (1976) 255--268.
    
\bibitem{zeidler}
    E.~Zeidler,
    Quantum Field Theory I. Basics in Mathematics and Physics.
    Springer, Berlin, 2006.
        
\end{thebibliography}
\end{document}